# Operando measurements of pressure-based PDMS deformations in electrochemical microfluidic devices

F. Lavanchy[1,2,3], Y. Sasaki[4,5], T. Minami[3,4], S. Chevalier[1,2,3*]

[1] Arts et Metiers Institute of Technology, CNRS, Bordeaux INP, I2M, UMR 5295, F-33400 Talence, France

[2] Univ. Bordeaux, CNRS, Bordeaux INP, I2M, UMR 5295, F-33400, Talence, France

[3] LIMMS/CNRS-IIS(UMI2820), The University of Tokyo, 4-6-1 Komaba, Meguro-ku, Tokyo, 153-8505, Japan

[4] Institute of Industrial Science, The University of Tokyo, 4-6-1 Komaba, Meguro-ku, Tokyo, 153-8505

[5] Research Center for Advanced Science and Technology, The University of Tokyo, 4-6-1, Komaba, Meguro-ku, Tokyo,153-8904 Japan

**Abstract**

Polymer based microfluidics, using polydimethylsiloxane (PDMS), are highly adaptable platforms for electrochemical studies commonly used for numerous applications. Deformations of these PDMS devices due to the flow pressure drop in microchannels have been widely documented in literature but cannot be prevented in all applications even though it is critical to measure quantitative data. In this work, we show that in electrochemical microfluidic devices used in energy conversion systems, the velocity profiles and channel deformations can be quantitatively measured based on imaging spectro-electrochemistry and numerical simulation of Hagen-Poiseuille flows. PDMS deformations are found amplified in the case of large channel aspect ratios (i.e. close to 100), with deformation $\Delta h/h_0 > 100\%$ at 50 kPa, and start to be non-negligible ($>

10%) for channel internal pressure above 5 kPa. Finally, by considering these deformations in the Beer-Lambert law, we show that accurate absorbance and concentration measurements can be done using visible spectroscopy regardless the channel internal pressure.

**Corresponding author:**

Prof. S. Chevalier
Mechanical Engineering Institute (I2M) of Bordeaux (CNRS UMR 5295)
Arts et Métiers Institute or Technology
Esplanade des Arts et Métiers
33405 Talence Cédex, France
Email: stephane.chevalier@u-bordeaux.fr

## 1. Introduction

Polymer-based microfluidic, in particular using polydimethylsiloxane (PDMS), became, since the introduction of lab-on-a-chip devices in the early 1990s, one of the main technologies to fabricate microchannels [1]. Although this technology significantly eases the fabrication process, the commonly used PDMS formulation (here with a 9:1 base-to-curing-agent ratio) exhibits pronounced hyper-elastic behavior. With a Young's modulus on the order of 1 MPa [2–4], this material remains highly deformable. To quantify this effect, several studies have demonstrated that PDMS microchannels undergo measurable deformation under known pressure conditions induced by imposed flow rates [5,6].. This deformation alters the channel cross-sectional geometry, leading to significant changes in fluid velocity and compromising the accuracy of spectroscopic measurements [2–4]. In Gervais et al. [5], it is demonstrated that in shallow channels (high aspect ratio, defined as $W/h$), pressure of few tens of kPa causes the channel walls to bulge. They identified a law where the channel height varies linearly with pressure as $h(p) = h_0 + c_1 Wp/E$, where $W$ is the channel width, $p$ is the pressure, $E$ is the Young modulus, and $c_1$ is a constant. Such law was then confirmed by Hardi et al. [6] using fluorescence microscopy to measure the PDMS channel deformations in the limit of $W/h \sim 20$. Several other studies investigated the effect of solvent-induced PDMS swelling [7]. Thus, although PDMS deformation is a widely known phenomenon, little is known when deformations are beyond $\Delta h/h$ >100%, and in the case of large channel aspect ratio ($W/h > 100$). In addition, the development of an operando imaging method to quantify these deformations, when they cannot be avoided, is highly desirable to predict and correct them in order to perform quantitative measurements in lab-on-chip devices.

Large channel aspect ratios are commonly encountered in microfluidic electrochemical devices (MED) [8–10]. These devices are used to convert electricity to chemical energy and, conversely, by controlling the electrochemical reaction at the electrode surface. They are used in a wide range of applications, described in the following reviews [11,12]. These devices are made in PDMS, but little has been reported about channel deformations in these technologies, although high flow rates are used to enhance the electrochemical reaction and avoid the reactants mixing in case of membraneless devices [13]. However, by taking advantage of the presence of electrodes in MED, it was demonstrated in previous studies that it is possible to have an operando access to the velocity profile across the microfluidic channels [14–19]. Thus, applying similar measurements to PDMS-based MED would enable a direct estimation of channel deformations, beyond 100%, in large channel aspect ratio, therefore completing the former studies reported by Gervais et al. and Hardi et al., and quantifying the limit of using high flow rates in PDMS-based MED.

The objective of this work is to show that reactant concentration oscillations in MED channels can be created using potential-step chronoamperometry with sinusoidal modulation or potential-modulated spectro-electrochemistry (or even with electrochemical impedance spectroscopy (EIS)) and used to measure the velocity profiles from the spatial phase shift imaged by a charge-coupled device (CCD) camera. Channel height deformations can then be estimated for a range of internal channel pressures from 10 kPa to 50 kPa and studied above 100% in the case of a large channel aspect ratio, set originally to 100 in this study. The methodology and mathematical framework used to make such measurements are detailed in the following sections. Finally, it is shown that, in the case PDMS deformations cannot be avoided, it is still possible to take them into account to perform quantitative measurements, such as light absorbance measurements, which rely on the channel heights.

## 2. Methods

### *1. Fabrication of the electrochemical microfluidic device*

An in-house microfluidic redox flow battery (RFB) was fabricated for imaging purposes. It comprises two Platinum (Pt) electrodes deposited on a Borofloat substrate (2'' wafer). The electrodes are 500 µm wide by 4.5 mm long and they are separated by 1 mm. A polydimethylsiloxane (PDMS with a 9:1 ratio base-curing agent, from SYLGARD 184 W/C - Japanese product name: SILPOT 184 W/C) T-channel covers the electrode to flow the reactants. The geometry of the microfluidic channel is 3 mm wide by 30 µm height, ensuring a large channel aspect ratio (i.e. $W/h \sim 100$ here). This particularity is twofold: (i) it reduces the light absorption in the height direction ensuring enough signal to noise ratio (SNR) to use transmission spectroscopy, and (ii) the velocity profile far from the wall (i.e. more than 50 µm is considered far in our case) follows the Hagen-Poiseuille law between two flat plates [20,21]. This later assumption holds in the central 90% of the channel width in our case. The top view of the cell is given in Figure 1(a), and a detailed description of the fabrication process can be found in the following reference [22].

Any fluids producing an electrochemical reaction on electrodes and having an absorbance peak in visible could have been chosen to perform this study. Here, the choice was made to use two aqueous solutions with at the anode 0.5 M of 4-hydroxy-2,2,6,6-tetramethylpiperidine 1-oxyl (4-HO-TEMPO) (Tokyo Chemical Industry, purity above 98%) plus 1.5 M of sodium chloride (NaCl) (Tokyo Chemical Industry, purity above 98%) dissolved in Milli-Q water. Similarly, at the cathode, a solution using 0.5 M of methyl viologen (MV) (Tokyo Chemical Industry) with 1.5 M of NaCl dissolved in Milli-Q water was prepared. Both solutions flow into the cell using a microfluidic pump (Fluigent Flow EZ) and were degassed continuously before entering the cell using Systec Mini Vacuum Degassing Chamber to remove the dissolved oxygen and keep the $MV^{+}$ ions stable after reacting. All the measurements are carried out at one flow rate, set at $Q_v = 5$ µl/min for each reactant (10 µl/min in total, corresponding to a theoretical velocity of 1.85 mm/s). The pressure is measured using the microfluidic pump, and the channel internal is controlled by increasing the pressure drop at the channel outlet. Electrochemical operating conditions were controlled using a potentiostat (Biologic SP-300) to measure the voltage and the current injected into the cell as well as its impedance using large amplitude sinusoidal voltage mode. Since no reference electrode was introduced in the channel, and due to the large current used in this study, it was decided to performed electrical measurements in a two electrodes configuration (without the use of reference electrode). Such configuration is accurate enough to generate the needed concentration oscillations in the channel (no electrochemical measurements are made in this study). Oxidation of 4-HO-TEMPO was performed at the working electrode.

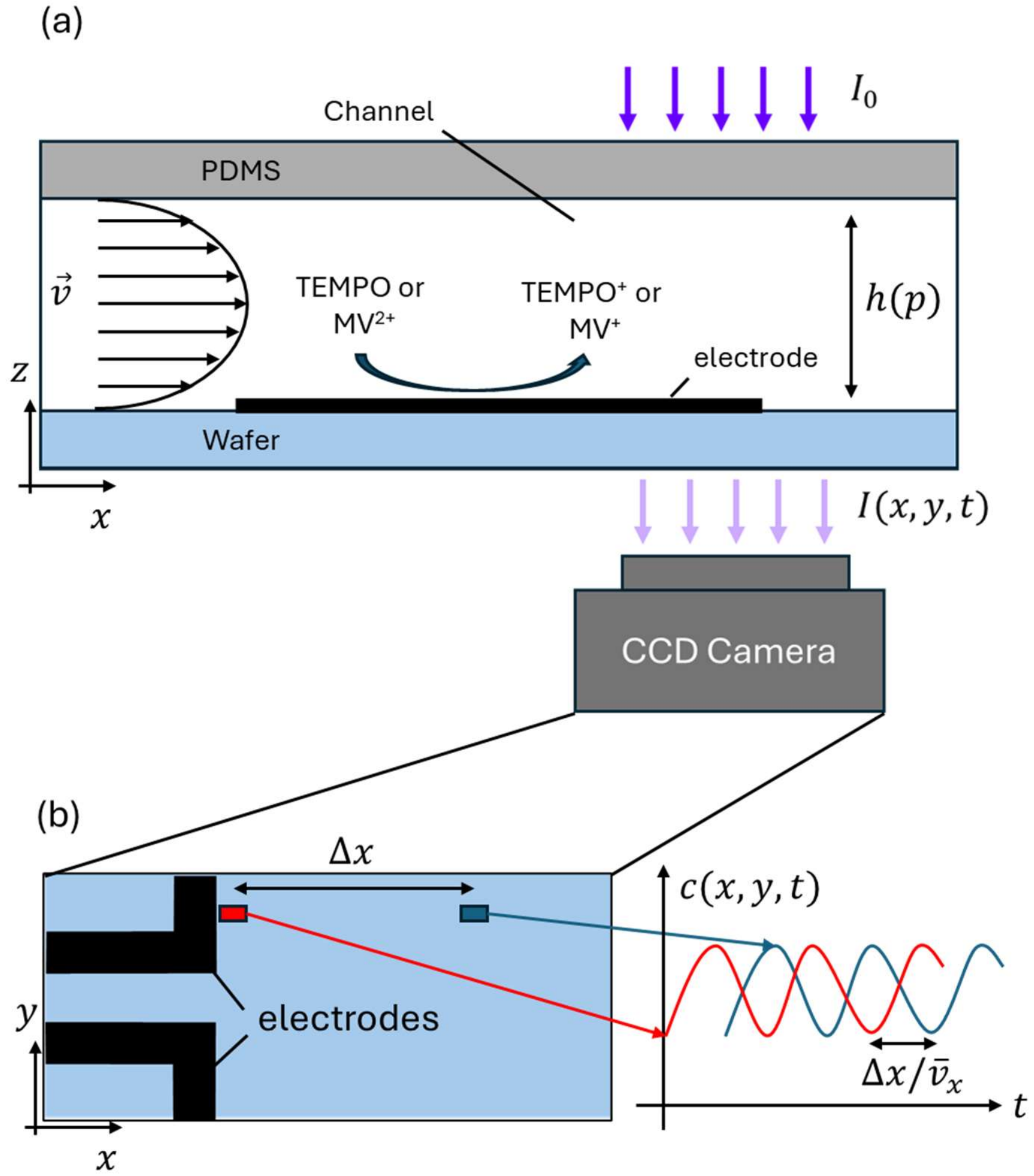

*Figure 1 - (a) Illustration of the microfluidic device and the electrochemical reaction taking place at the electrode surface, (b) The imaging method and the induced concentration oscillations in the microchannel. The phase shift between two points in the channel is proportional to the average fluid velocity.*

Images were collected through an Olympus IX70 inverse microscope. A schematic of our imaging setup is presented in Figure 1(a). The primary light source was a halogen white lamp with a collimation adapter placed around 10 cm above the cell. A narrow bandpass filter ($\lambda = 435 \pm 5$ nm from custom-made Edmund Optics PN956351) was used to produce a monochromatic blue light passing through the RFB. This wavelength corresponds to the absorption peak of -TEMPO [23] and the lowest absorption of MV in the visible domain (see the supplementary Figure S2). Since the absorbance of MV is 100 times higher than TEMPO, the SNR is high enough to image both compounds at the same time. The light was finally collected through an 8-bit CCD camera (Infinity 2-1R Lumenera). The resulting spatial resolution is 1.16 µm/px, leading to a field of view of approximately 1.21 mm by 1.62 mm. It is important to note that in this work, the microscope is used as spectrophotometers to measure the light absorbance and molar concentration of compounds. A camera with higher bit depth could have been used to improve the SNR, however, in this study it was sufficient to measure the chemical absorbances of both reactants.

### 2. *Measurement of the velocity profile*

In this section, we describe the measurements of the actual fluid velocity using imaging spectro-electrochemistry [22]. The methodology is based on creating TEMPO and MV concentration oscillations downstream to the electrodes in the microfluidic channels [14,16] (see Figure 1(b)). By doing so, it is known that the phase shift between two positions in the channel is directly linked to the local velocity (see Figure 1(b)).

Concentration oscillations are created using a sinusoidal potential with a frequency $f = 0.25$ Hz or 0.5 Hz, and an amplitude of $\Delta E_0 = 100$ mV around a baseline value $\overline{E} = 1.3$ V as $E(t) = \overline{E} + \Delta E_0 \cos(\omega t)$, where $\omega = 2\pi f$ is the angular frequency. Each measurement lasts for a duration of $t_{max} = 40$ s corresponding to a recording of 10 or 20 periods.

To model the concentration oscillation field, Fick's law is expressed in 3D (assuming no source term, $S = 0$, as we consider the flow downstream to the electrodes in the channel) as follows:

$$\frac{\partial c}{\partial t} + \vec{v} \cdot \vec{\nabla} c - D\nabla^2 c = 0 \tag{1}$$

where $c$ is the molar concentration, D is the diffusivity (here the diffusivity is considered constant inside the channel), $\vec{v}$ is the fluid velocity vector, $t$ the time, $x$ the length of the channel and $y$ the width of the channels. In the case of large channel aspect ratio, as described earlier, the concentration can be considered uniform in z-direction and equation (1) can be written in 2D as

$$\frac{\partial c}{\partial t} + \bar{v}_x(y)\,\frac{\partial c}{\partial x} - D\left(\frac{\partial^2 c}{\partial x^2} + \frac{\partial^2 c}{\partial y^2}\right) = 0, \tag{2}$$

with $\bar{v}_x(y)$ the fluid velocity in the x-direction [20], defined as the average velocity along z-axis at one y position (as $\overline{v}_x(y) = \int_0^{h(y)} v_x(z,y)/h(y)dz$). By assuming that Peclet number is $Pe = \overline{v}_x L/D \sim 55$, equation (2) can be written as:

$$\frac{1}{\overline{v}_x(y)}\frac{\partial c}{\partial t} + \frac{\partial c}{\partial x} - \frac{L}{Pe}\left(\frac{\partial^2 c}{\partial x^2} + \frac{\partial^2 c}{\partial y^2}\right) = 0, \tag{3}$$

and considering that $Pe >> 1$, it yields to $\overline{v}_x(y)\frac{\partial c}{\partial x} \gg \frac{\partial^2 c}{\partial x^2}$ and $\frac{\partial^2 c}{\partial y^2}$. In such high velocity conditions, Taylor-Aris regime arises and increases the diffusion in x-direction [24] (estimated to be $D_{eff} \sim 10D$ in our conditions) while homogenizing the concentration in z-direction. However, the effective diffusion in x and y-direction is largely negligible compared to the advection transport $\overline{v}_x(y)\frac{\partial c}{\partial x}$, except close to the wall at $y = 0$ and $y = l_c$. Thus, the transport equation in our microchannel can be well described using the following equation:

$$\frac{\partial c}{\partial t} + \overline{v}_x(y)\frac{\partial c}{\partial x} = 0. \tag{4}$$

When the concentration is modulated the electrode outlet (i.e. $c(x = 0, y = 0, t) = \Delta c\cos(\omega t)$), a solution to this equation can be found as [15,25]

$$c(x, y, t) = \overline{c} + \Delta c\cos(\omega t + \varphi(y, x)), \tag{5}$$

where $\varphi(x, y) = -\,\omega x/\,\overline{v}_x(y)$ is the phase which depends on the position x downstream to the electrode and the local velocity $\overline{v}_x(y)$. Therefore, by measuring the phase at each position x and using a linear regression, we can access the local velocity. By repeating this process for each single y position in the width of the channel one can obtain the velocity profile $\overline{v}_x(y)$.

*3. Image processing*

Measurement of the concentration and the associated phase downstream to the electrodes can be done using the Beer-Lambert law. Thus, the absorbance A is defined as follows:

$$A(x, y, t) = -\log\left(\frac{I(x, y, t)}{I_0}\right) = \kappa\, h(y)\, c(x, y, t), \tag{6}$$

where $I$ and $I_0$ represent the transmitted light intensity through the sample and the incident light intensity in the absence of any absorbing medium, respectively. Moreover, $\kappa$ is the molar absorption coefficient (or molar extinction coefficient [26]). In the case of sinusoidal steady-state response of a linear time-invariant (LTI) system state [27] we can write:

$$-\log\frac{\overline{I} + \Delta I\cos(\omega t + \varphi_{abs}(x, y))}{I_0} = \kappa h(y)\,(\,\overline{c} + \Delta c\cos(\omega t + \varphi(x, y))) \tag{7}$$

To move forward, one can factorize by $\overline{I}/_{I_0}$,assuming that we have the term $\Delta I/_{\overline{I}} \ll 1$, and with a Taylor expansion of order 1 around 0 we obtain:

$$-\log\left(\frac{\bar{I}}{I_0}\right) - \frac{\Delta I}{\bar{I}}\cos\big(\omega t + \varphi_{abs}(x,y)\big) \; = \; \kappa h(y)\,(\,\bar{c} + \Delta c\cos(\omega t + \varphi(x,y))) \tag{8}$$

By terms identification, we can write:

$$\begin{cases} \kappa\, h(y)\Delta c = \; -\dfrac{\Delta I}{\bar{I}} \\ \varphi_c(x,y) = \varphi_{abs}(x,y) = -\,\dfrac{\omega x}{\bar{v}_x(y)} \end{cases} \tag{9}$$

Measurements of the phase $\varphi_{abs}$ in every camera pixel (see Figure 1) is done by using the temporal Fourier transform of the images. For each pixel, the following calculation is performed:

$$\hat{I}(x,y,\omega) = \int_0^{t_{max}} I(x,y,t)e^{-i\omega t}dt, \tag{10}$$

where $\hat{I}$ is the Fourier transform of the signal $I$ measured by the camera (see Figure 1). Finally, the phase of the absorbance is given by its argument as:

$$\varphi_{abs}(x,y) = \arg\left(\hat{I}(x,y,\omega)\right) = -\frac{\omega}{\bar{v}_x(y)}x, \tag{11}$$

where arg is the complex argument. Therefore, for every y-position, the velocity can be estimated from the linear regression of $\varphi_{abs}(x,y)$ along the x-coordinate of the image.

## 3. Results

### *1. Velocity field measurements*

The cell was operated at 1.3V with a 100 mV oscillation superimposed around the baseline. The camera records a 40 s video from which the phase signal is extracted using a Fourier transform.

The oscillation frequency is set at 0.25 Hz at high pressure and 0.50 Hz at low pressure to improve the SNR. Typical phase fields obtained downstream to the electrodes are given in Figure 2(b) and (c) for TEMPO and MV, respectively. As expected, the phase shift increases in the flow direction. At each y-position, the phase also varies, as it depends on the local fluid velocity, which is not uniform across the channel [20]. By comparing Figure 2(b) and (c), it can be observed that the signal recorded using MV is of better quality compared to TEMPO, since the absorptivity coefficient of TEMPO is 100 times lower than MV [23,26]. However, to address this issue, we applied an averaging in y direction with a step size of 20 pixels, followed by a 20 pixels moving average in the x direction. This preprocessing allowed us to smooth the data and improve the visibility of the underlying concentration gradients.

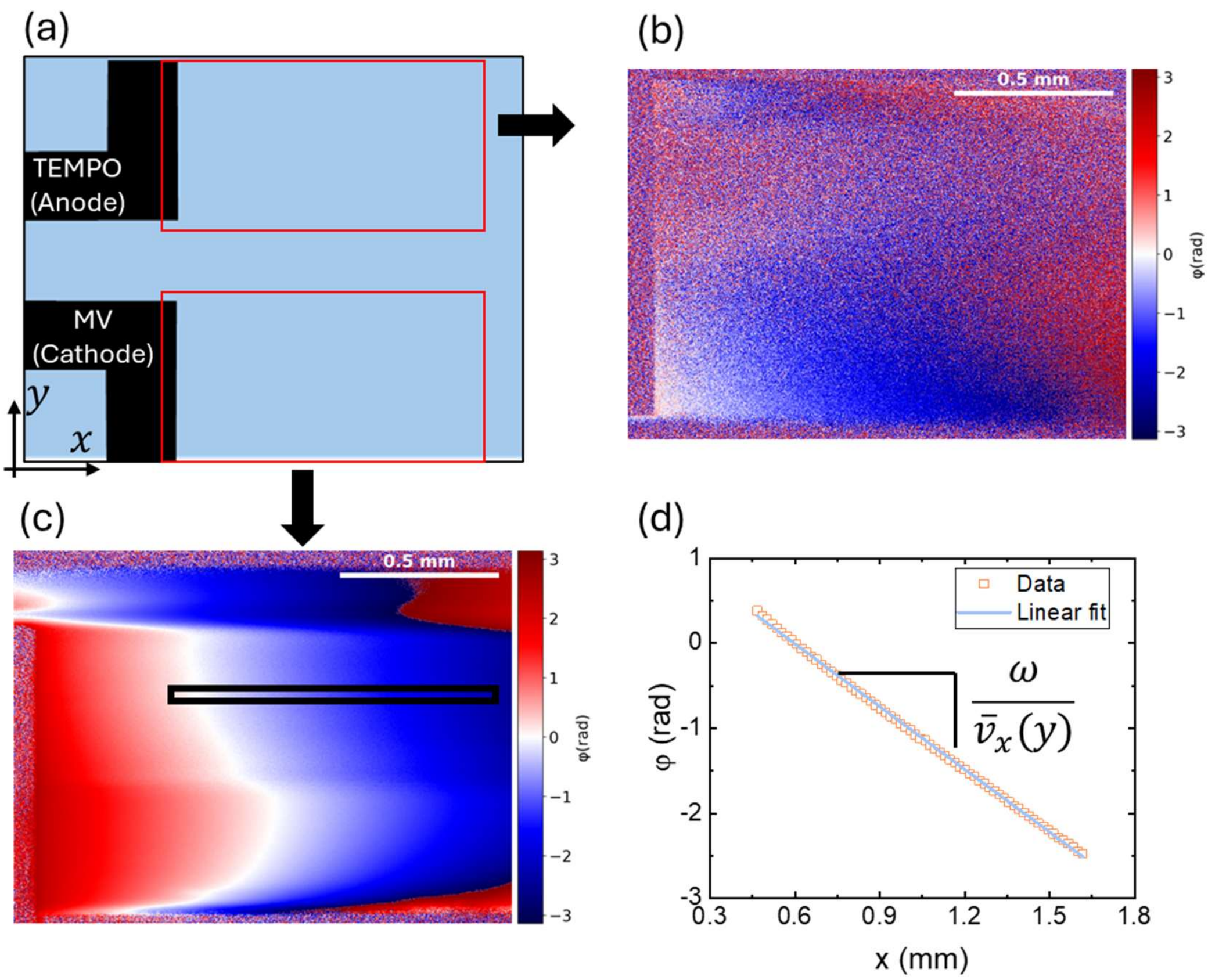

*Figure 2 - (a) Schematic of the chip in x-y plan and its region of interest, (b) Phase field of TEMPO at 33 kPa, (c) Phase field of MV at 33 kPa, (d) Measurements of the local velocity by linear regression for a fixed y (depicted in the black rectangle in (c)).*

In Figure 2(d), a typical linear regression is given for one y position from which the velocity is extracted. By repeating the same linear regression through all the y positions, the velocity field $\overline{v}_x(y)$ in the channel width is measured. The same phase field measurements and velocity field estimations were repeated for 4 channel internal pressures (made by increasing the pressure drop at the outlet).

*2. PDMS deformation as function of pressure*

The phase field images obtained were converted into velocity profiles through the channel width. The measurements are presented in Figure 3(a) for the four-pressure used at a total fixed flow rate of 10 µl/min (5 µl/min of each reactant). It is observed that the velocity profiles increase from the channel walls (at $y = \pm 1.5$ mm) toward the middle of the channels. In theory, the average velocity defined as $\bar{v} = 2Q_v \,/\, (h_0 W)$ should be 1.85 mm/s; however, such a velocity is never reached, even at the lowest pressure. The magnitude of velocity profiles keeps decreasing as pressure increases, indicating a significant deformation of the PDMS top wall.

Both the velocity profiles measured from the TEMPO and MV phase field have led to the same velocity profiles within $\pm 5$ % uncertainty. However, the phase fields from TEMPO concentration oscillation are particularly challenging to process, as the absorbance of TEMPO at 435 nm is approximately 100 times lower than that of MV [23,26]. This significantly reduces the SNR, making the data more susceptible to noise. Additionally, the presence of phase jumps must also be considered, further complicating the analysis when the SRN is low. As a result, the precision of the TEMPO measurements is lower than that of MV.

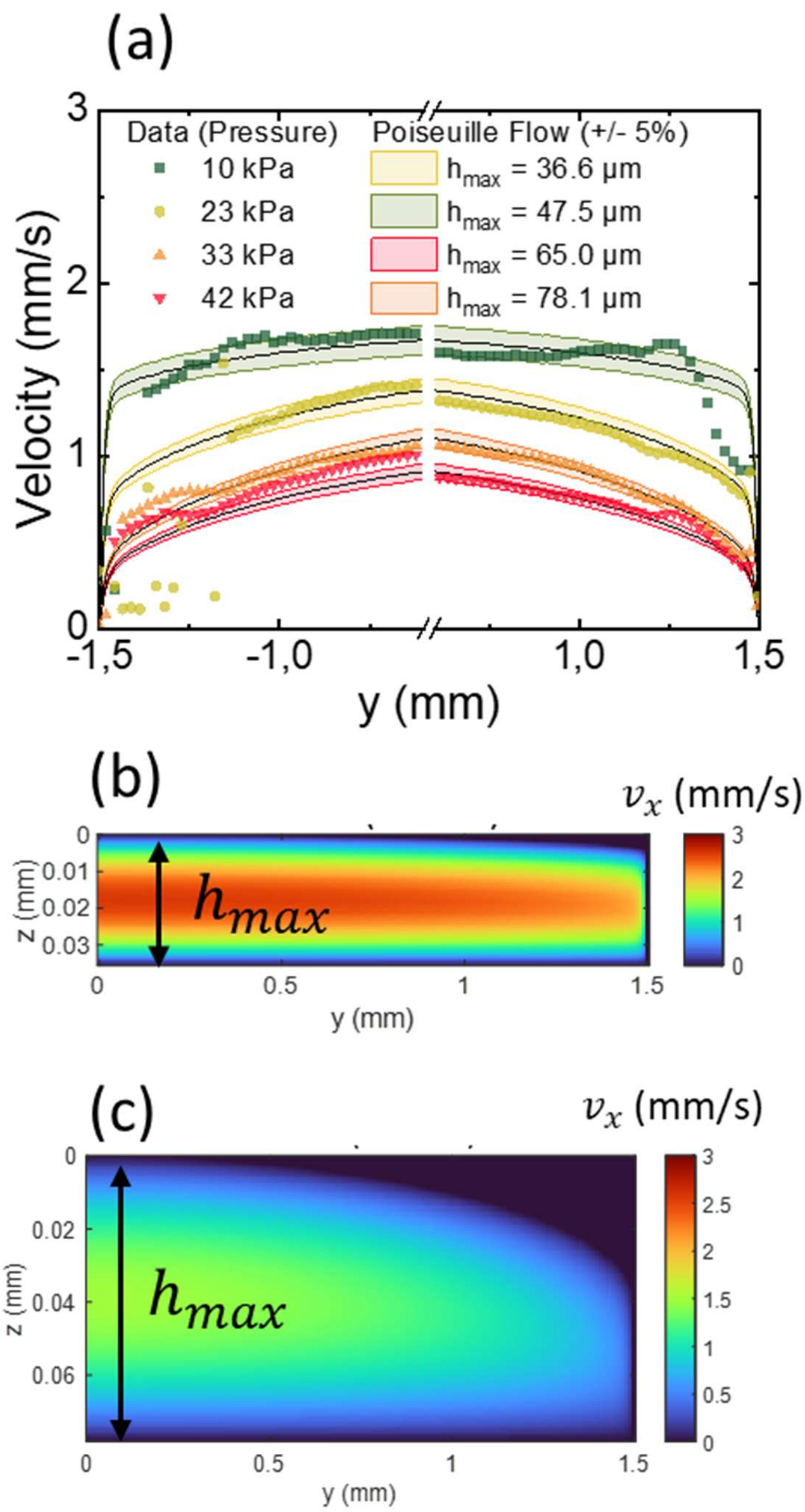

*Figure 3 – (a). Fluid velocity fields obtained for a range of channel internal pressure and comparison with the computed Poiseuille Flow in deformed PDMS: (b) 10 kPa and (c) 42 kPa.*

To explain the shape of the velocity profiles measured and estimate the maximum channels heights, theoretical Hagen-Poiseuille flows were computed in deformed PDMS channels. Top wall elliptical shape was assumed as reported previously in literature [6,28], and the following equation was used

$$\frac{\partial^2 v_x}{\partial y^2} + \frac{\partial^2 v_x}{\partial z^2} = \frac{\Delta p}{\mu L} \tag{11}$$

where $\mu$ is the fluid dynamic viscosity, $L$ the channel length and $\Delta p$ the inlet and outlet pressure difference (different from the total pressure measured in the microchannel). The pressure drop was set to ensure a constant flow rate of 5 µl/min (regardless of the channel height), and no slip boundary conditions were used on walls. The calculation was done only on half channel given the symmetry of the profiles observed, and the MATLAB® PDE toolbox was used to solve equation 11 using finite element method. The simulations were repeated for a range of maximum channel height until they fit the velocity profiles measured. These results are given in Figure 3(a) to (c).

The numerical velocity profiles in Figure 3(a) agree well with the measurements within the $\pm 5$ % interval represented by the colored surface. The shape of the velocity profile is mainly due to the PDMS deformations (see the maps in Figure 3(b) and (c)). The average velocity profile in a rigid rectangle channel would have been flat after few hundred of µm far from the side walls [20,21]. Therefore, PDMS deformations change both the magnitude and velocity distribution which can be really damaging for optimal control of the operating conditions.

From the maximum channel height obtained numerically, the channel deformation $\Delta h/h_0$ was computed and presented versus the channel internal pressure in Figure 4. As predicted in former studies, a linear deformation is globally observed in our data. However, at deformation beyond 100%, the behavior seems not linear anymore and could be fitted by polynomial laws. Such observations are confirmed by the traction curves measured on our PDMS samples (see the supplementary Figure S1), where the nonlinear domain is found to start above 100% deformations.

If we consider only the linear regime, the slope was found to be $(2.9 \pm 0.5) \times 10^{-2}$ $kPa^{-1}$. According to Gervais et al. [28] the slope is $\propto c_1 W/(Eh)$ where $E \approx 1.1$ MPa and $W = 3$ mm (see Figure S1 in supplementary material). In our case, we obtain $c_1$ around $0.30 \pm 0.05$ which is compatible with the data reported in literature for such a large channel aspect ratio, validating our measurements of local velocity profiles, the assumption made, and extended the range of validity of PDMS deformation to a range greater than 100% using a large channel aspect ratio ($W/h > 100$). For Hele-Shaw cells, which share the same mechanical configuration as our device, with a ratio $W/h \gg 1$, our operando method could, in principle, be applied to such systems. However, in order to observe and quantify the flow velocity within a Hele-Shaw cell, it is necessary to generate a measurable sinusoidal signal that can be detected by the camera. This temporal modulation is essential, as it enables velocity measurements at a fixed spatial location and by repeating this measurement at multiple spatial positions along the channel length, it would then be possible to reconstruct and image the two-dimensional deformation of the Hele-Shaw cell.

Last but not least, it is important to note that our results show that significant PDMS deformations (higher than 10%) occur at relatively low pressure (below 5 kPa), which can be easily achieved by flowing aqueous solutions at a few tens of µl/min. In our 10 mm long channel with 10 µl/min flow rate, the pressure drop can be estimated to be $\Delta p = 12\mu L Q_v/(Wh^3) \approx 2.5$ kPa where a deformation greater than 10 % can already be observed.

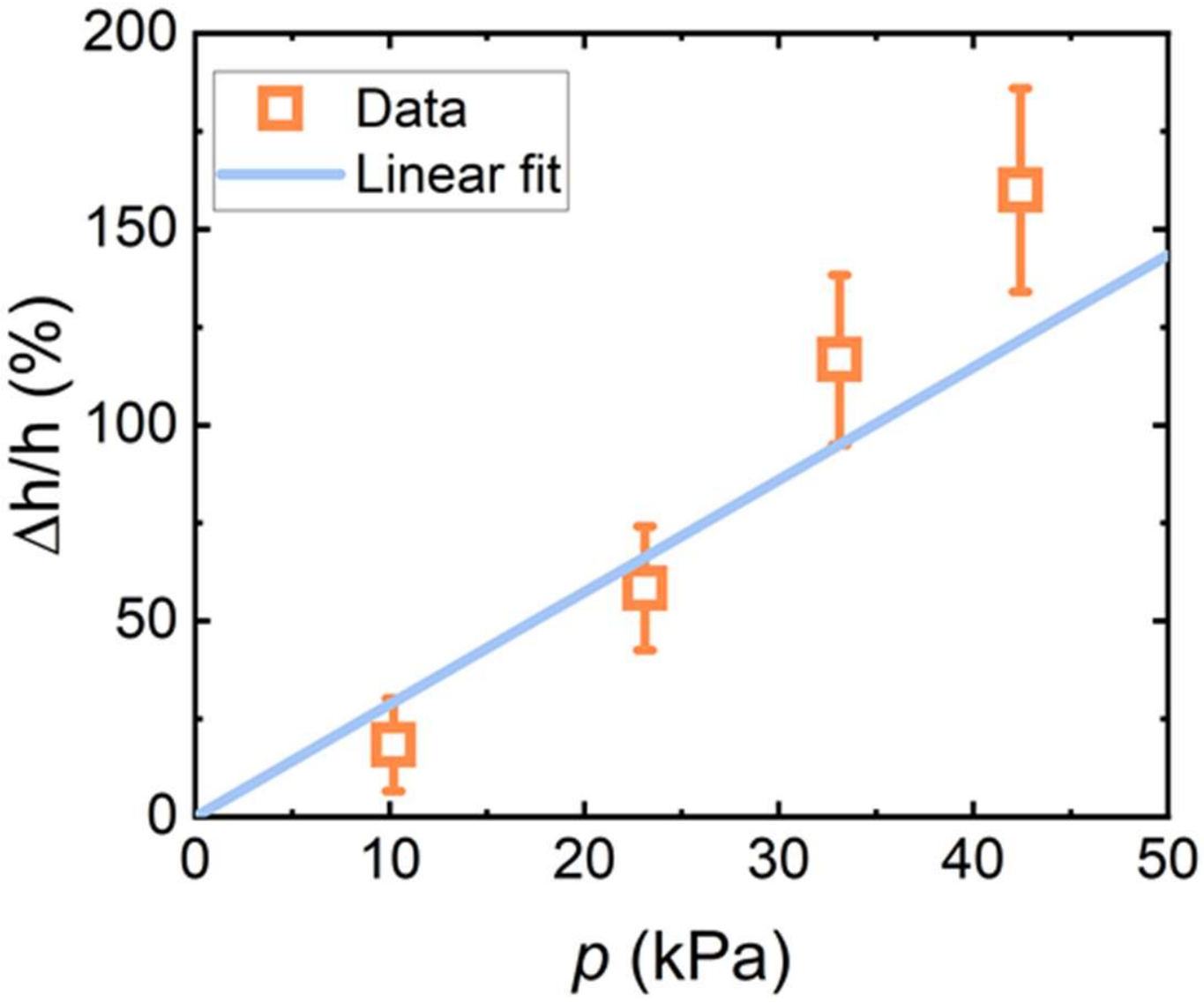

*Figure 4 - Maximum deformation estimated in the channel as function in channel internal pressure, $R^2$=0.95*

### 3. *Absorbance measurements under PDMS deformation*

To complete this study and demonstrate that, if PDMS deformations cannot be avoided, quantitative results can still be obtained if the channel heights are re-estimated, absorbance measured at different flow rates and channel internal pressure were performed using the same imaging setup. The absorbance coefficients of MV and TEMPO are estimated at 435 nm. In MED, the link between the current injected in the cell and the variation of reactant concentration is given by the Faraday law [19] as:

$$j = zFQ_{v_eq}\left(\int_0^l c(x,y)dy - \int_0^l c_{OCV}(x,y)dy\right), \tag{12}$$

where $j$ is the current injected in the electrochemical device, $Q_{v_eq} = 2lQ_v/W$, $l$ is the image width, i.e. 1.21 mm. Then, one can link the current injected to the cell (in steady state) to the absorbance given by the Beer-Lambert law as

$$j = \frac{zFQ_{v_eq}}{\kappa h(P)}\left(\int_0^l A(x,y)dy - \int_0^l A_{OCV}(x,y)dy\right) = \frac{zF}{\kappa}\frac{Q_{v_eq}}{h(P)}\Delta A, \tag{13}$$

where $\Delta A = \int_0^l \log(I(x,y)/I_{OCV}(x,y))\,dy$, $I_{OCV}(x,y)$ is the pixel signal measured during OCV [22]. So, a linear relationship is found between the current injected into the cell, the relative absorbance measured and the channel height which varies according to the pressure in the microchannel. Therefore, in the case where the PDMS deformations are not taking into account, the linearity of equation 13 should be broken.

As it can be seen in Figure 5(a) and (b), a clear linear relationship has been observed regardless of the pressure or flow rate used, validating our measurement and the use of our methods to make quantitative measurement even in presence of PDMS deformations. The linear slope is determined for each reactant, and the absorptivity coefficient can be obtained as $\kappa_{TEMPO} = zF/38400 \approx 2.51 \pm 0.02$ $M^{-1}.mm^{-1}$ and $\kappa_{MV} = zF/310 \approx 311 \pm 5$ $M^{-1}.mm^{-1}$ confirming that $\kappa_{\mathrm{MV}} \approx 124\kappa_{\mathrm{TEMPO}}$. These values are, to the authors' best knowledge, the first operando measurement of MV and TEMPO absorptivity measurements at these wavelengths, even in the presence of channel deformations.

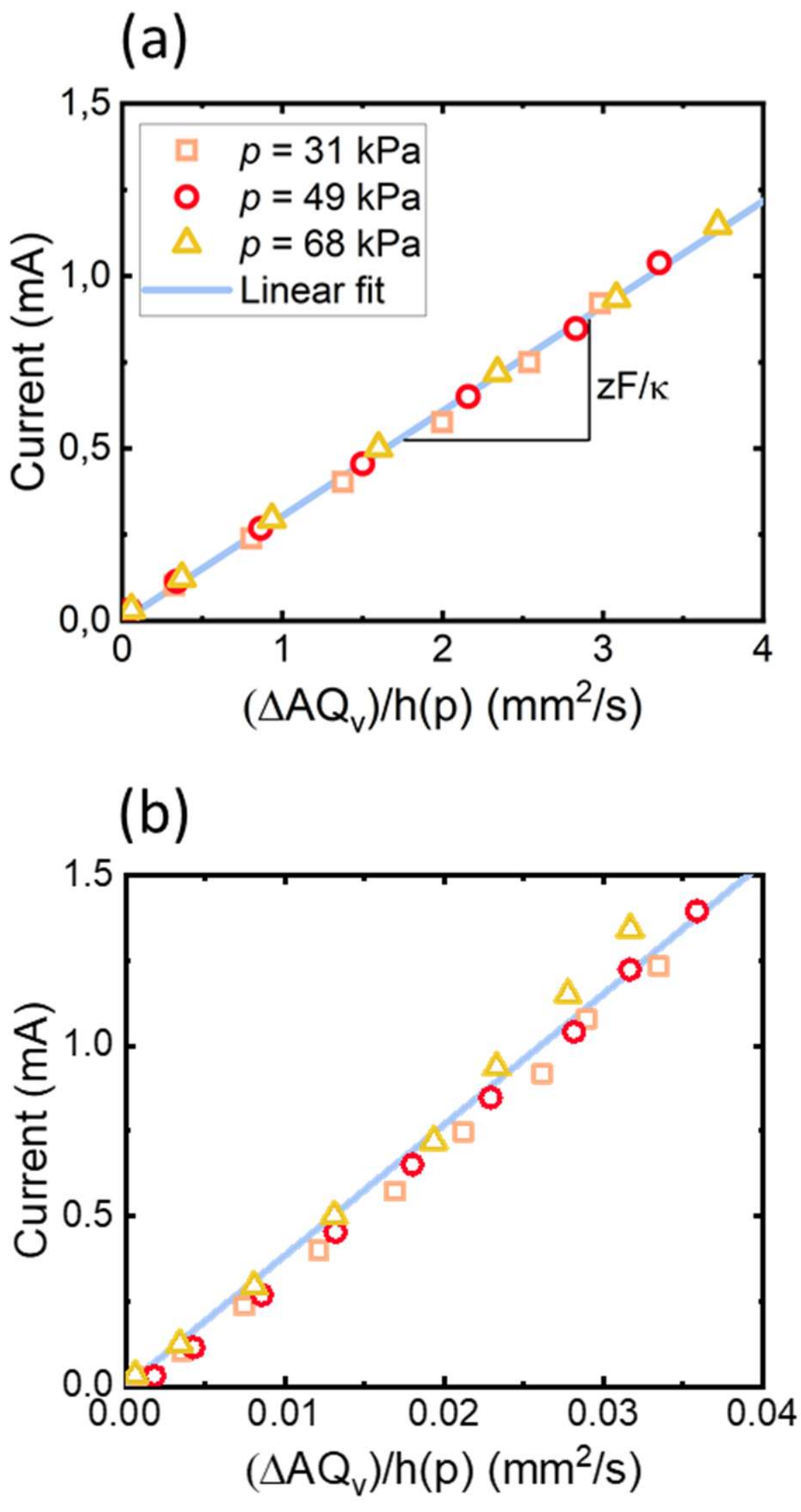

*Figure 5-Measurement of the absorbance coefficient for a range of flow rate, current and pressure (so channel height). (a) for MV and (b) for TEMPO.*

## 4. Conclusion

In work, a novel methodology based on concentration oscillations in the microfluidic channel was presented and used to measure operando channel deformations up to 150% channel height deformation (in a large aspect ratio device). The method used takes advantage of the concentration oscillation naturally created downstream to the electrodes during EIS measurements and enables us to estimate the velocity profile and channel height. Our results show that the linear law between channel height deformation and channel internal pressure was correctly measured up to 100% deformation, but beyond, nonlinear behavior starts to be observed. It is also important to note that these deformations, because of the large channel aspect ratio, start to be significant even for pressure as low as few kPa (corresponding to the typical pressure drop obtained with flow rate of few tens of µl/min in our 30 µm by 3 mm channel)

Thus, one of the main conclusions of this study is that since high flow rates are usually used through PDMS-based microfluidic device to ensure high kinetic rates, channel deformations are very likely to appear. However, if a measurable concentration oscillation can be induced, these channel deformations can be taken into account during quantitative measurements, such as operando concentration imaging based on spectroscopic techniques, as demonstrated by our measurements of MV and TEMPO molar absorptivity coefficients. It should, however, be noted that the large difference in absorbance between TEMPO and MV (TEMPO being approximately 100 times less absorbing than MV at the selected wavelength) introduces significant noise in TEMPO-based measurements, thereby calling for the development of more suitable spectroscopic or analytical approaches to obtain sharper and more robust velocity profiles.

More broadly, this work lays the foundations for operando metrology of deformable microfluidic devices, enabling mechanical effects long regarded as parasitic to be transformed into measurable and exploitable parameters, with direct perspectives for high-throughput screening and the quantitative study of complex reaction systems under strong hydrodynamic constraints. Beyond microfluidics, this method could also be envisioned for measuring flow velocities in blood vessels, which undergo significant deformation under pressure.

**Acknowledgments**

The authors gratefully acknowledge the CNRS and the University of Tokyo through the Joint PhD Program (SmartBat) for funding this work. Dr. Shintaro Nakagawa and Prof. Naoko Yoshie of the University of Tokyo are also gratefully acknowledged for their kind help during the traction curve measurements.

## References

[1] H. Becker, *Polymer microfluidic devices*, Talanta. **56** (2002) 267–287.

[2] H.M. Xia, J.W. Wu, J.J. Zheng, J. Zhang, Z.P. Wang, *Nonlinear microfluidics: device physics, functions, and applications*, Lab Chip. **21** (2021) 1241–1268.

[3] H. Fallahi, J. Zhang, H.-P. Phan, N.-T. Nguyen, *Flexible Microfluidics: Fundamentals, Recent Developments, and Applications*, Micromachines (Basel). **10** (2019) 830.

[4] Y. Yalikun, N. Ota, B. Guo, T. Tang, Y. Zhou, C. Lei, H. Kobayashi, Y. Hosokawa, M. Li, H. Enrique Muñoz, D. Di Carlo, K. Goda, Y. Tanaka, *Effects of Flow-Induced Microfluidic Chip Wall Deformation on Imaging Flow Cytometry*, Cytometry Part A. **97** (2020) 909–920.

[5] T. Gervais, K.F. Jensen, *Mass transport and surface reactions in microfluidic systems*, Chem. Eng. Sci. **61** (2006) 1102–1121.

[6] B.S. Hardy, K. Uechi, J. Zhen, H. Pirouz Kavehpour, *The deformation of flexible PDMS microchannels under a pressure driven flow*, Lab Chip. **9** (2009) 935–938.

[7] R. Dangla, F. Gallaire, C.N. Baroud, *Microchannel deformations due to solvent-induced PDMS swelling*, Lab Chip. **10** (2010) 2972.

[8] O.A. Ibrahim, M. Goulet, E. Kjeang, *In-situ characterization of symmetric dual-pass architecture of microfluidic co-laminar flow cells*, Electrochim. Acta. **187** (2016) 277–285.

[9] E. Kjeang, A.G. Brolo, D.A. Harrington, N. Djilali, D. Sinton, *Hydrogen Peroxide as an Oxidant for Microfluidic Fuel Cells*, J. Electrochem. Soc. **154** (2007) B1220.

[10] S.H. Lee, Y. Ahn, *A laminar flow-based single stack of flow-over planar microfluidic fuel cells*, J. Power Sources. **351** (2017) 67–73.

[11] M.A. Modestino, D. Fernandez Rivas, S.M.H. Hashemi, J.G.E. Gardeniers, D. Psaltis, *The potential for microfluidics in electrochemical energy systems*, Energy Environ. Sci. **9** (2016) 3381–3391.

[12] O.A. Ibrahim, M. Navarro-Segarra, P. Sadeghi, N. Sabaté, J.P. Esquivel, E. Kjeang, *Microfluidics for Electrochemical Energy Conversion*, Chem. Rev. **122** (2022) 7236–7266.

[13] S. Chevalier, Y. Sasaki, T. Minami, *Image-based measurements of Tafel slopes in aqueous MV/4-HO-TEMPO Flow Batteries*, J. Power Sources. **655** (2025) 237928.

[14] I.A. Schneider, S.A. Freunberger, D. Kramer, A. Wokaun, G.G. Scherer, *Oscillations in Gas Channels. Part I. The Forgotten Player in Impedance Spectroscopy in PEFCs*, J. Electrochem. Soc. **154** (2007) B383–B388.

[15] D. Kramer, I.A. Schneider, A. Wokaun, G.G. Scherer, *Oscillations in the Gas Channels - The Forgotten Player in Impedance Spectroscopy in Polymer Electrolyte Fuel Cells B. Modeling the Wave*, in: ECS Trans., ECS, 2006: pp. 1249–1258.

[16] A. Kulikovsky, O. Shamardina, *A Model for PEM Fuel Cell Impedance: Oxygen Flow in the Channel Triggers Spatial and Frequency Oscillations of the Local Impedance*, J. Electrochem. Soc. **162** (2015) F1068–F1077.

[17] S. Chevalier, C. Josset, A. Bazylak, B. Auvity, *Measurements of Air Velocities in Polymer Electrolyte Membrane Fuel Cell Channels Using Electrochemical Impedance Spectroscopy*, J. Electrochem. Soc. **163** (2016) F816–F823.

[18] S. Chevalier, J.-C. Olivier, C. Josset, B. Auvity, *Polymer electrolyte membrane fuel cell operating in stoichiometric regime*, J. Power Sources. **440** (2019) 227100.

[19] E. Kjeang, B. Roesch, J. McKechnie, D.A. Harrington, N. Djilali, D. Sinton, *Integrated electrochemical velocimetry for microfluidic devices*, Microfluid. Nanofluidics. **3** (2007) 403–416.

[20] H. Bruus, *Theoretical microfluidics*, Oxford university press, 2008.

[21] S. Chevalier, *Semianalytical modeling of the mass transfer in microfluidic electrochemical chips*, Phys. Rev. E. **104** (2021) 035110.

[22] M. Garcia, A. Sommier, D. Michau, J.-C. Batsale, S. Chevalier, *Operando Spectroelectrochemical Imaging of Concentration Fields and Tafel Kinetics in Microfluidic Electrochemical Devices*, Anal. Chem. **96** (2024) 16487–16492.

[23] K. Żamojć, M. Zdrowowicz, W. Wiczk, D. Jacewicz, L. Chmurzyński, *Dihydroxycoumarins as highly selective fluorescent probes for the fast detection of 4-hydroxy-TEMPO in aqueous solution*, RSC Adv. **5** (2015) 63807–63812.

[24] P. Tabeling, *Introduction to microfluidics*, 2nd ed., 2023.

[25] S. Chevalier, C. Josset, B. Auvity, *Analytical solution for the low frequency polymer electrolyte membrane fuel cell impedance*, J. Power Sources. **407** (2018) 123–131.

[26] N. Aristov, A. Habekost, *Electrochromism of Methylviologen (Paraquat)*, World Journal of Chemical Education. **3** (2015) 82–86.

[27] L. Zadeh, C. Desoer, *Linear system theory: the state space approach*, Courier Dover Publications, 2008.

[28] T. Gervais, J. El-Ali, A. Günther, K.F. Jensen, *Flow-induced deformation of shallow microfluidic channels*, Lab Chip. **6** (2006) 500.